\documentclass[aps, prl, superscriptaddress, amsmath, amssymb, reprint]{revtex4-1}

\pdfoutput=1

\usepackage{graphicx}
\usepackage{hyperref}

\begin{document}

\title[Demonstration of the synchrotron type spectrum of Betatron
radiation]{Demonstration of the synchrotron type spectrum of Laser-produced
Betatron radiation}

\author{S.~Fourmaux}
\email{fourmaux@emt.inrs.ca}
\affiliation{INRS-EMT, Universit\'e du Qu\'ebec, 1650 Lionel Boulet, Varennes
J3X 1S2, Qu\'ebec, Canada}
\author{S. Corde}
\author{K. Ta Phuoc}
\affiliation{Laboratoire d'Optique Appliqu\'ee, ENSTA ParisTech -
CNRS UMR7639 - \'Ecole Polytechnique, Chemin de la Huni\`ere,
Palaiseau F-91761, France}
\author{P. M. Leguay}
\author{S.~Payeur}
\author{P.~Lassonde}
\author{S.~Gnedyuk}
\author{G.~Lebrun}
\affiliation{INRS-EMT, Universit\'e du Qu\'ebec, 1650 Lionel Boulet, Varennes
J3X 1S2, Qu\'ebec, Canada}
\author{C.~Fourment}
\affiliation{Centre Lasers Intenses et Applications (CELIA), Universit\'e de
Bordeaux-CNRS-CEA, Talence F-33405, France}
\author{V.~Malka}
\author{S.~Sebban}
\author{A.~Rousse}
\affiliation{Laboratoire d'Optique Appliqu\'ee, ENSTA ParisTech -
CNRS UMR7639 - \'Ecole Polytechnique, Chemin de la Huni\`ere,
Palaiseau F-91761, France}
\author{J.~C.~Kieffer}
\affiliation{INRS-EMT, Universit\'e du Qu\'ebec, 1650 Lionel Boulet, Varennes
J3X 1S2, Qu\'ebec, Canada}


\begin{abstract}
Betatron X-ray radiation in laser-plasma accelerators is produced when electrons are accelerated and wiggled in the laser-wakefield cavity. This femtosecond source, producing intense X-ray beams in the multi
kiloelectronvolt range has been observed at different interaction regime using
high power laser from 10 to 100 TW. However, none of the spectral measurement
performed were at sufficient resolution, bandwidth and signal to noise ratio to precisely determine the
shape of spectra with a single laser shot in order to avoid shot to shot fluctuations. In this letter, the Betatron radiation produced using a 80 TW
laser is characterized by using a single photon counting method. We measure in single shot
spectra from 8 to 21 keV with a resolution better than 350 eV. The results
obtained are in excellent agreement with theoretical predictions and demonstrate
the synchrotron type nature of this radiation mechanism. The critical energy
is found to be $E_c=5.6\pm1\:\textrm{keV}$ for our experimental conditions. In
addition, the features of the source at this energy range open novel
perspectives for applications in time-resolved X-ray science.
\end{abstract}

\maketitle

A femtosecond X-ray beam, called Betatron, can be produced by focusing an
intense femtosecond laser pulse at relativistic
intensities, on the order of $10^{18}-10^{19}\:\textrm{W.cm}^{-2}$,  onto a gas jet target. Interacting
with the quasi-instantaneously created under-dense plasma, the laser pulse
excites a wakefield in which electrons can be trapped and accelerated to
high energies in short distances \cite{PRL1979Tajima, APB2002Pukhov,
Nature2004Geddes, Nature2004Mangles, Nature2004Faure}. These electrons perform
Betatron oscillations across the propagation axis, and emit X-ray photons
\cite{PRE2002Esarey, PoP2003Kostyukov, PRL2004Kiselev, PRL2004Rousse,
PRL2006TaPhuoc} (radiation from accelerating charged-particles). The Betatron
radiation consists on a broadband X-ray beam, collimated within 10's mrad, with
a femtosecond duration \cite{PoP2007TaPhuoc}.

During the past few years, several experiments have been dedicated, at different
laser facilities, to the characterization of Betatron radiation. Even if
the origin of the radiation was clearly identified, its spectrum has never been
precisely determined. This information is however crucial to improve our knowledge of
the physical mechanisms driving the source, identify the electrons participating to the emission, and determine the most appropriate routes for its development.  In addition, for any potential application the precise shape of the spectrum must be known.

So far, spectra estimations were either based
on the measurement of the transmission through an array of filters or by using the diffraction from crystals.
The use of filters is the most elementary method and it allows a single shot measurement. The results obtained using this method are generally fitted with the synchrotron distribution theoretically
predicted \cite{PRL2008Kneip, PROC2009Kneip, APL2009Mangles, NatPhys2010Kneip}. However, this rely on the assumption that the spectrum is synchrotron-like and can not give any
deviation from such distribution, or details in the structure of the spectrum.
When the Bragg diffraction from a crystal is used, the resolution is important but the
characterization range is limited to about 1-3 keV \cite{PRE2008Albert} and the measurement requires an accumulation over about 10 laser shots for each energy point. Consequently this method is very sensitive to the strong fluctuations of the Betatron spectrum and can only provide a mean spectrum of the source.
To overcome the limitations of the precedent methods, the photon counting can be a relevant method. For a sufficiently intense source and an appropriate experimental setup, we will show that it can provide a single shot measurement of the source over a large bandwidth.
A photon counting based measurement of the Betatron source has been recently used in the range 1-9 keV and a continuous spectrum was observed, but its structure was not revealed since it was not deconvoluted by the filters transmission and the CCD response \cite{RSI2010Thorn, PROC2010Plateau}.

In this letter, we present single shot photon counting measurements of the Betatron X-ray radiation spectrum in the 8-21 keV energy range with a resolution better than 350 eV. Thanks to this method, the results  demonstrate the synchrotron type nature of the Betatron radiation and its direct correlation with the accelerated electron energy spectrum which were simultaneously measured. In the experiment presented, the Betatron radiation was produced at the interaction of a 80 TW / 30 fs laser pulse with a gas jet target density on the order of $10^{18}-10^{19}\:\textrm{cm}^{-3}$. We will show that the experimental spectrum fits a synchrotron distribution of critical energy $E_c=5.6\pm1\:\textrm{keV}$.

In a laser-plasma accelerator, electrons are both accelerated longitudinally and
wiggled transversally by the electromagnetic wakefields. The transverse
oscillation is nearly sinusoidal at the Betatron frequency
$\omega_\beta=\omega_p/\sqrt{2\gamma}$ \cite{PRE2002Esarey, PoP2003Kostyukov},
where $\gamma$ is the relativistic factor of the electron and
$\omega_p=(n_ee^2/m_e\epsilon_0)^{1/2}$ the plasma frequency, with $n_e$ the
electron density, $e$ the electron charge and $m_e$ the electron mass. Due to
this oscillatory motion, radiation is emitted with properties depending on the
strength parameter $K=r_\beta k_p\sqrt{\gamma/2}$ ($r_\beta$ is the Betatron
transverse amplitude of motion and $k_p=\omega_p/c$), on the Betatron frequency
$\omega_\beta$ and on the electron energy $\mathcal{E}=\gamma mc^2$. For
$K\ll1$, the on-axis radiation spectrum is nearly monochromatic at the
fundamental frequency $\omega=2\gamma^2\omega_\beta/(1+K^2/2)$. As
$K\rightarrow1$, harmonics of the fundamental start to appear in the spectrum,
and for $K\gg1$ the spectrum contains many closely spaced harmonics and extends
up to a critical frequency $\omega_c=3K\gamma^2\omega_\beta/2$. Experimental
data \cite{PRL2004Rousse, PRL2006TaPhuoc, PRL2008Kneip, PROC2009Kneip,
APL2009Mangles} have shown that Betatron oscillations occur in the wiggler
regime, $K\gg1$. In this latter regime, for a single electron with constant
parameters $K$, $\omega_\beta$ and $\gamma$, the integrated radiation spectrum
is similar to the synchrotron one which is given by $dI/d\omega\propto
S(x=\omega/\omega_c)=\int_x^\infty K_{5/3}(\xi)d\xi$, where $K_{5/3}$ is a
modified Bessel function of the second kind. But because electrons are
accelerating with changing parameters and because different electrons may have
different parameters, a Betatron spectrum is probably more correctly described
by a sum of synchrotron distributions \cite{PoP2010Thomas}.

\begin{figure}
   \centering
   \includegraphics[angle=-90, width=8.5cm]{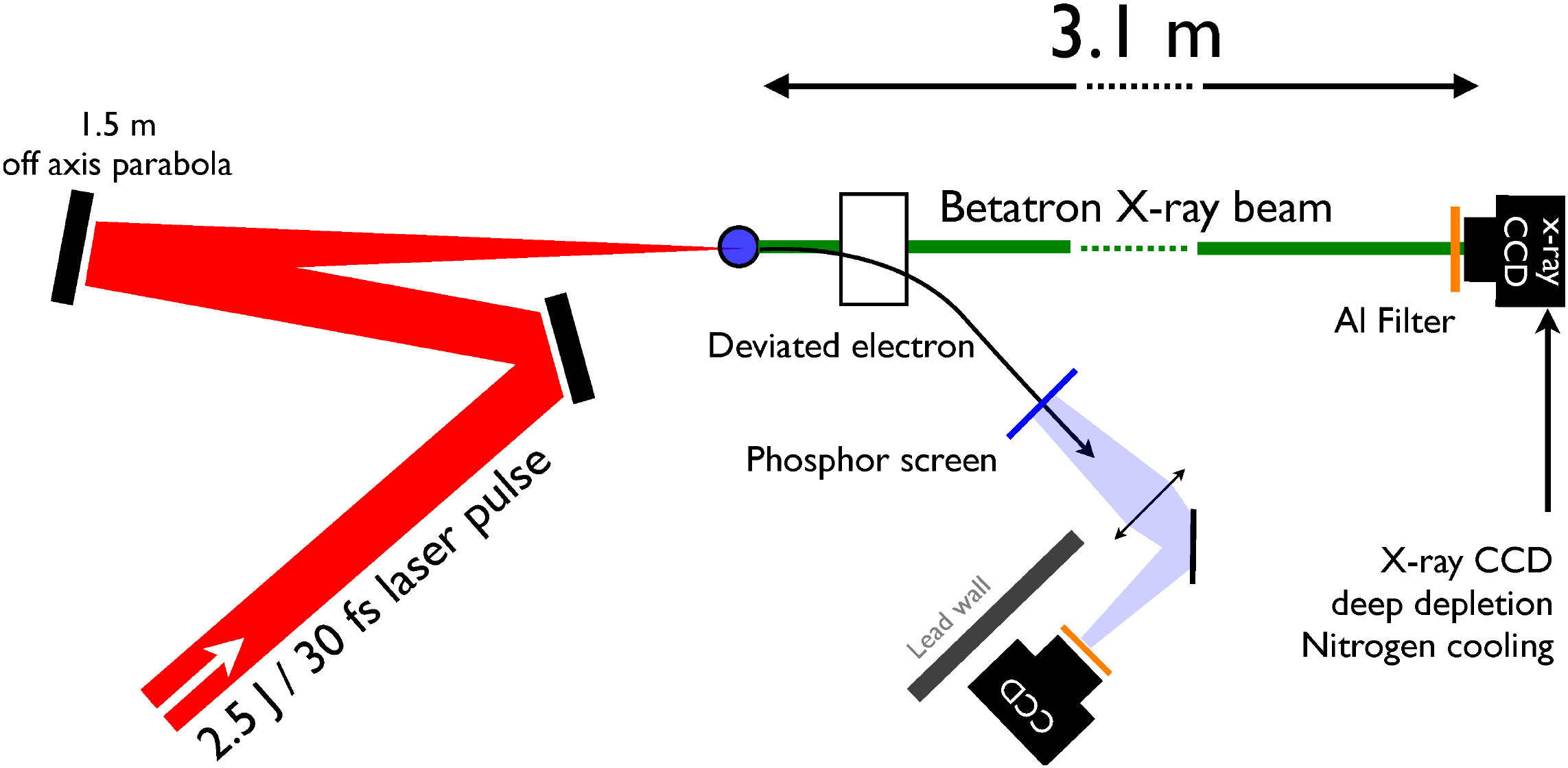}
   \caption{Schematic of the experimental set-up for electron acceleration and
Betatron X-ray generation.}
   \label{fig1}
\end{figure}

The experiment has been performed at the Advanced Laser Light Source (ALLS)
facility at INRS-EMT \cite{OE2008Fourmaux}, using a titanium-doped sapphire
(Ti:sapphire) laser operating at 10 Hz with a central wavelength of 800 nm in
chirped-pulse amplification mode. The ALLS laser system delivered 2.5 Joule of
energy on target with a full width at half maximum (FWHM) duration of 30 fs
(80TW) and linear polarization. The experimental set-up for electron
acceleration and Betatron X-ray generation is presented on figure \ref{fig1}.
The laser pulse was focused by an $f=1.5$ m off-axis parabolic mirror onto a supersonic helium gas jet. In the focal plane, the FWHM spot size was 24 $\mu\textrm{m}$ and the encircled energy in this FWHM spot size was 30\% of the total energy. It corresponds to an initial laser intensity of $3\times10^{18}\:\textrm{W.cm}^{-2}$ and a normalized vector potential amplitude of $a_0=eA_0/m_ec=1.2$. The gas jet density profile has been characterized by interferometry \cite{RSI2000Malka}. In this experiment, we used a 3 mm diameter helium gas jet whose density profile has a well-defined 2.1-mm-long electron density plateau of $n_e=5.4\times10^{18}\:\textrm{cm}^{-3}$.

\begin{figure}
   \centering
   \includegraphics[angle=-90, width=8.5cm]{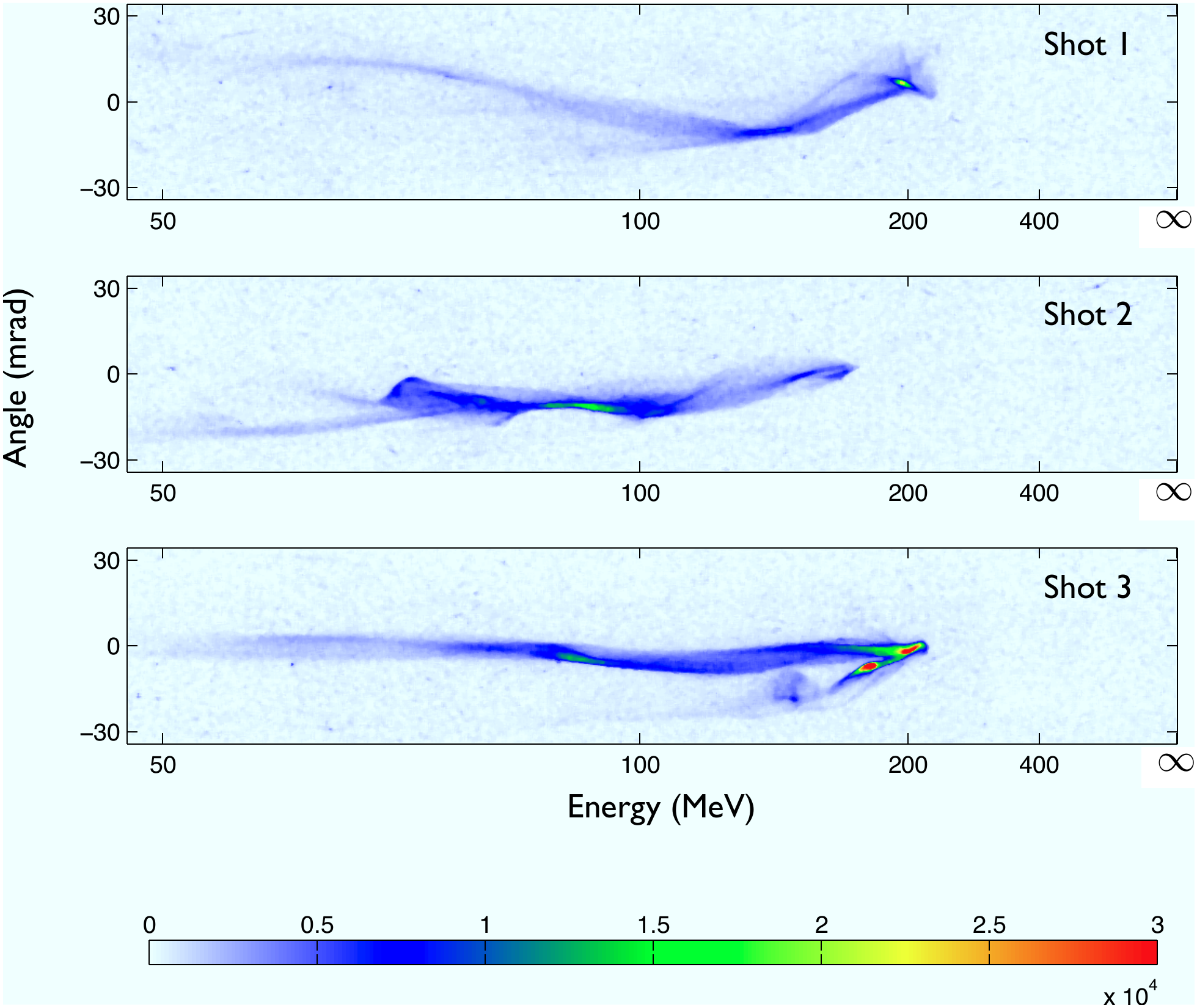}
   \caption{Three typical raw electron spectra (shot 1, 2, and 3). Horizontal axis, electron
energy; vertical axis, exit angle; color scale, number of counts. This latter
gives an indication of the beam charge.}
   \label{fig2}
\end{figure}

The electron beam produced in the interaction is measured with a spectrometer
consisting of a permanent dipole magnet (1.1 T over 10 cm) which deflects
electrons depending on their energy, and a Lanex phosphor screen to convert a
fraction of the electron energy into 546 nm light imaged by a CCD camera \cite{RSI2006Glinec}. Three
typical raw electron spectra recorded in the experiment are displayed on figure
\ref{fig2}. 
It shows that electrons are accelerated up to approximately 200 MeV, and that transverse structure are present in the raw spectra (correlation between energy and exit angle), which are reminiscent of the Betatron motion in the laser-plasma accelerator \cite{EPL2008Glinec}.

X-rays produced by the accelerated electrons have been measured using a deep-depletion X-ray CCD (model PI-LCX:1300 cooled with liquid nitrogen), with $1340\times1300$ imaging pixels of size $20\mu$m$\times 20\mu$m. The X-ray CCD was directly connected and vacuum pumped via the interaction chamber. The quantum efficiency extends well above 20 keV, allowing to count X-ray photons beyond ten keV.
First, we placed the X-ray CCD close to the X-ray source (distance of 1.2 m) in order to measure the X-ray angular profile. Typical measured angular spreads of the X-ray beams were on the order of 20 mrad (FWHM). Using an array of aluminum filters of different thicknesses (4, 34, 64 and 124 $\mu$m), the measurement of the transmission through each filter can be used to fit for the best synchrotron distribution reproducing the data. Using the synchrotron distribution $S(\omega/\omega_c)$ defined above, we obtained a best fit for $E_c=\hbar\omega_c=5.7\:\textrm{keV}$. However, this method is very imprecise and does not allow to obtain any detail in the X-ray spectrum, or any deviation from a unique synchrotron distribution.

A precise measurement of the X-ray spectrum can be achieved by photon counting \cite{RSI2008Maddox, RSI2009Fourment}.
The CCD camera is composed of $1340\times1300$ pixels, i.e. 1,740,000
independent detectors, and a single photon detected by one of these detectors
gives a number of counts (analog to digital converter (ADC) unit) which is
proportional to its energy: $N_c=\alpha \hbar\omega$, where $N_c$ is the number
of counts and $\hbar\omega$ the photon energy. For our ADC settings, we obtained
$\alpha=0.11$ count per eV by calibrating the X-ray CCD using $K_\alpha$ lines
emitted in laser-solid interaction and using the Betatron X-ray beam passing
through a Cu filter which has a sharp cut-off at 8.98 keV. If the number of
photons per pixel is small compared to one, piling events (several photons
detected on a single pixel) can be neglected and the measurement of the X-ray
spectrum becomes possible. A single photon leads to the formation of an electron
cloud in the Silicon layer of the CCD chip, which can spread over several
neighbor pixels. This phenomenon has to be taken into account in the data
analysis. We have used a first algorithm able to detect events spreading over a
few pixels (multi-pixel event, MPE) and a second algorithm which only takes into
account non-spreading events in which the electron cloud is detected on only one
pixel (single-pixel event, SPE). For a MPE, the photon energy can be recovered
by summing the number of counts over all pixels of the event. However, we found
that the MPE algorithm had a lower energy resolution and was more sensitive to
piling events than the SPE algorithm. On the other hand, to recover the
experimental spectrum from the SPE method, it is necessary to know the
probability that a single photon yields a single-pixel event, $k1(\hbar\omega)$,
which depends on the photon energy $\hbar\omega$. This function has been
obtained from a simulation modeling our CCD response and providing
$k1(\hbar\omega)$ \cite{RSI2009Fourment}.
The experimental spectrum is then recovered by:
\begin{equation}
\frac{dN_X}{d(\hbar\omega)}=\frac{a}{k1(\hbar\omega)QE(\hbar\omega)T(\hbar\omega)}\frac{dN_c}{d(\hbar\omega)}\times\frac{dN_{SPE}}{dN_c},
\end{equation}
where $dN_{SPE}/dN_c$ is the number of single-pixel events for each number of counts obtained from the SPE algorithm, $dN_c/d(\hbar\omega)=\alpha=0.11$ count/eV, $QE(\hbar\omega)$ is the CCD quantum efficiency, $T(\hbar\omega)$ is the transmission of the filters placed before the CCD, $a$ is a numerical factor and $dN_X/d(\hbar\omega)$ is the number of photons per unit energy (in eV$^{-1}$). The numerical factor $a$ comes from the fact that many SPE photons are not analyzed by the algorithm because they are superposed or situated next to other photons. This factor $a$ is obtained by requiring that the spectrum $dN_X/d(\hbar\omega)$ leads to the correct total number of counts in the CCD image.
It should be noted that contrary to reference \cite{RSI2010Thorn}, the studied energy range doesn't contain any edge or line emission making the analysis easier. In our work, the energy range is limited at low energy by the Al filter thickness used to reduce the number of photons on the CCD camera. The energy range could be extended to lower energies by setting the detector farther away.

In order to operate in photon counting mode, we placed the X-ray CCD at a distance of 3.1 m from the source as shown on figure
\ref{fig1}, and we attenuated the signal using an Al filter with a thickness of
274 $\mu$m. We collected photons in a solid angle of $\Omega=7.3\times10^{-5}$
sr around the propagation axis.

\begin{figure}[h]
   \centering
   \includegraphics[width=8.5cm]{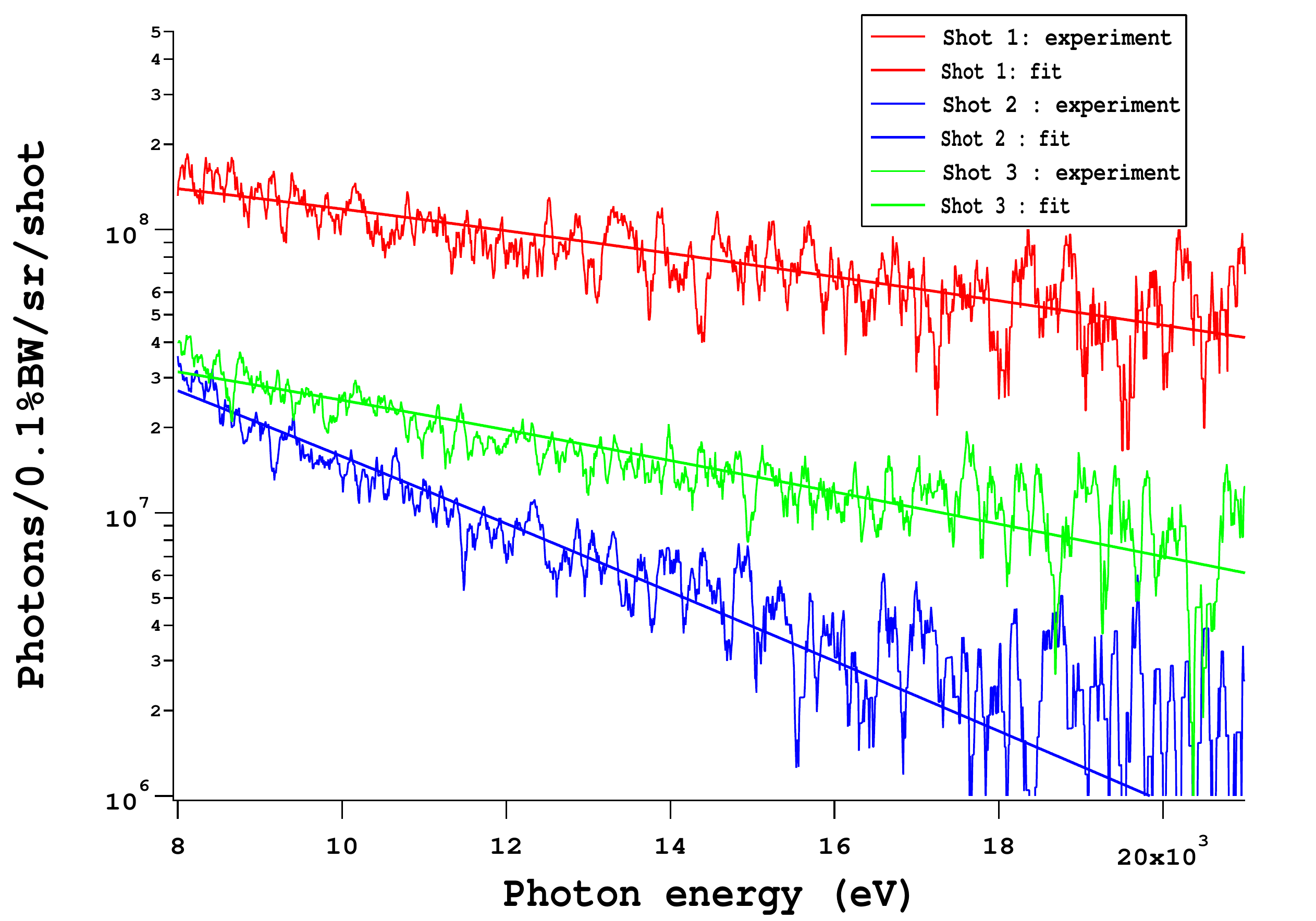}
   \caption{Spectra of Betatron X-rays obtained from photon counting, for laser shots 1 (red), 2 (green), and 3 (blue) corresponding to the electron spectra given on figure 2, and the best fit with a synchrotron distribution for each laser shot.}
   \label{fig3}
\end{figure} 
Figure \ref{fig3} displays the measured experimental spectra of Betatron X-rays by photon counting corresponding to each laser shot whose electron energy spectra have already been shown on figure 2. The estimated energy resolution is better than 350 eV. A fit of the experimental measurements by a synchrotron distribution (of the type $S(\omega/\omega_c)$ as defined above) is also shown for each laser shot. The best fit was respectively obtained with a synchrotron distribution of critical energy $E_c=\hbar\omega_c=8.5\:\textrm{keV}$, $E_c=\hbar\omega_c=3.2\:\textrm{keV}$, and $E_c=\hbar\omega_c=6.6\:\textrm{keV}$ (shot 1, 2, and 3).

Since both the electron and X-ray photon spectra are simultaneously obtained in a single laser shot they can be correlated. If we consider the shot 1 shown both on figure 2 and 3, we observe  important Betatron oscillations combined with a high electron charge at 200 MeV. This is well correlated with a high critical energy ($E_c=\hbar\omega_c=8.5\:\textrm{keV}$) and a high number of photons (more than $10^8$ photons/0.1\%bandwidth/sr/shot). For shot 2, the maximum electron charge is well below 200 MeV and the Betatron oscillations are small compared to shot 1. This is well correlated with the small critical energy $E_c=\hbar\omega_c=3.2\:\textrm{keV}$.  

\begin{figure}
   \centering
   \includegraphics[width=8.5cm]{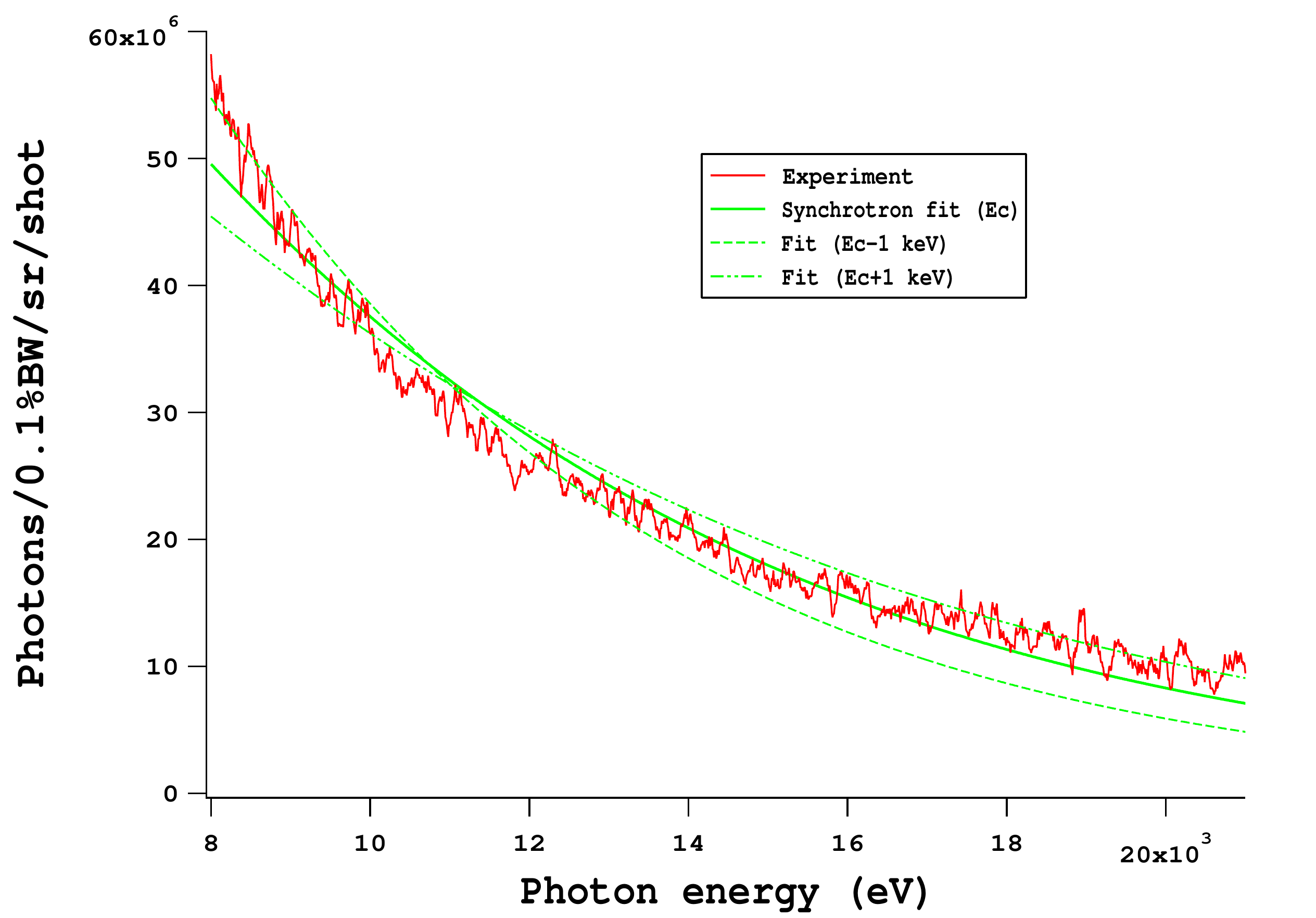}
   \caption{Spectrum of Betatron X-rays obtained from photon counting, averaged
over ten shots (in red line), and the best fit with a synchrotron distribution
of critical energy $E_c=\hbar\omega_c=5.6\:\textrm{keV}$ (in green line). We illustrate the precision over the critical energy determination by showing the synchrotron distribution corresponding to $E_c=\hbar\omega_c=5.6\pm 1\:\textrm{keV}$ (in green dashed lines).}
   \label{fig4}
\end{figure} 
Figure \ref{fig4} displays the measured experimental spectrum of Betatron X-rays averaged over ten successive shots. As the X-ray beam has a pointing fluctuation on the order of the beam size, it can be regarded as an averaged spectrum over angles. The average also allows to give a typical spectrum (since shot to shot fluctuations are important) and to improve the signal to noise ratio.  A fit of the experimental measurement by a synchrotron distribution (of the type $S(\omega/\omega_c)$ as defined above) is also shown. The best fit was obtained with a synchrotron distribution of critical energy $E_c=\hbar\omega_c=5.6\:\textrm{keV}$. The measurement precision over the critical energy is $\pm 1\:\textrm{keV}$.

In conclusion, we have presented a single shot and large spectral range characterization of laser-produced Betatron radiation. At this experiment, the source produces $3.6\times10^7$ and $1.1\times10^7$ photons/0.1\%bandwidth/sr/shot at respectively 10 and 20 keV. The result shows unambiguously that the single shot experimental spectra fit synchrotron distributions. The averaged spectrum has a best fit for $E_c=5.6\pm1\:\textrm{keV}$. The high critical energy obtained in these experiments demonstrate the potential of this X-ray source for diffraction and imaging applications. It also shows the interest for 100 TW scale laser system to go beyond 10 Hz repetition rate to increase the X-ray source average brightness.

\bigskip
The authors would like to thanks ALLS technical team for their support: M. O.
Bussi\`ere, J. Maltais, C. Morissette, L. Pelletier, F. Poitras, P. L. Renault
and C. Sirois. The ALLS facility has been funded by the Canadian Foundation for
Innovation (CFI). This work is funded by NSERC, the Canada Research Chair
program and Minist\`ere de l'\'Education du Qu\'ebec. We acknowledge the Agence
Nationale pour la Recherche, through the COKER project ANR-06-BLAN-0123-01.


\bibliography{sebastiencorde}

\end{document}